\documentclass[aps,twocolumn]{revtex4}
\usepackage{graphicx,subfigure}
\usepackage{amsmath,amssymb}
\usepackage{bm}

\newcommand{\be}{\begin{eqnarray}}
\newcommand{\ee}{\end{eqnarray}}

\def\slashchar#1{\setbox0=\hbox{$#1$}           
   \dimen0=\wd0                                 
   \setbox1=\hbox{/} \dimen1=\wd1               
  \ifdim\dimen0>\dimen1                        
 \rlap{\hbox to \dimen0{\hfil/\hfil}}      
  #1                                        
 \else                                        
    \rlap{\hbox to \dimen1{\hfil$#1$\hfil}}   
    /                                         
 \fi}                                         %

\begin{document}

\title{Comment on measurement of the rotaion frequency and the magnetic field \\
at the  freezeout of heavy ion collisions}

\author{  Edward  Shuryak }

\affiliation{Department of Physics and Astronomy, Stony Brook University,
Stony Brook NY 11794-3800, USA}

\begin{abstract}
We suggest other decays which can more effectively identify the effect of rotational motion and
presence of the magnetic field, compared to $\Lambda,\bar\Lambda$ hyperon decayse used so far.
\end{abstract}
\maketitle
Both (i) rotation of the fireball and (ii) the presence of the magnetic field at greezeout stage of
heavy ion collisions has been discussed before.
In particular, while partial capture of the initial strong pulse of magnetic field by
electric currents in QGP has been discussed by Tuchin\cite{Tuchin:2013apa}
 and others,
yet there is no consensus on possible magnitude of the  field at frezzeout.

The first measurements which produced non-zero results have
recently been reportted by STAR collaboration at RHIC \cite{STAR_pol}
using self-analysing weak decays of $\Lambda$ and $\bar\Lambda$ hyperons
and the fact that reaction plane direction for each event
is (with certain resolution) known.
Let us mention few important qualitative features of their data:\\
(a) The average magnitude of the effect is at $P_{av}=2-3 $\% level, increasing toward smaller collisions energy\\
(b) All data points show systematically that the effect is noticeably 
larger for  $\bar\Lambda$ than for $\Lambda$, by about  $\Delta P\approx 1\%.$

Let us take as a working hypothesis for further discussion that the average effect (a) is cause by 
rotation and the difference (b) by the magnetic field, reflecting the opposite charge and magnetic moment
of  $\bar\Lambda$ .  (There can of course be other detector-related reasons for such a difference,
which we for now would ignore.)

This leads to estimates of the rotational frequency $\omega$ and bagnetic field $B$
\be \omega= 2T tanh^{-1} (P_{av})\sim {1 \over 25 \, fm}\ee
\be eB= \Delta P {T \over  \mu_\Lambda }\approx 0.004 \, GeV^2\ee
 where we used the numbers mentioned above,
 the  usual $(\vec {\omega} \vec {J})$  rotational energy and $J_\Lambda=1/2$. 
 
 Our suggestion is that one can use wider variety of the decays to measure both contributions.
 
 Measurements of the  {\em magnetic field } can be better done with particles 
 posessing larger magnetic moment, than Lambda hyperons. We propose for this role $\Delta^{++}$
 resonance: its magnetic moment
 \be \mu_{ \Delta^{++}} =3\mu_u= 5.56\mu_N \ee
 is  nearly an order of magnitude larger than 
  \be \mu_{ \Lambda} \approx \mu_s=-0.61 \mu_N \ee
 and so, potentially, the polarization can perhaps be at the level of 10\%,
 if our working hypothesis is true. If not, it will provide the upper limit on $B$ at freezeout, which
 of great interest by itself.
 
 $Rotation$ can be more conveniently measured by decays of hadrons with $zero$ magnetic moments,
 such as e.g.
 \be \rho^0 \rightarrow \pi^+ \pi^- \ee
  \be \Delta^0 \rightarrow p \pi^- \ee
  which also have  $J=1,3/2$, two and three times larger than $J_\Lambda=1/2$. 

 In particular, the last decay should be compared to  \be \Delta^{++}\rightarrow p \pi^+ \ee
 we considered above as the one most sensitive to magnetic field.
 Unlike the $\bar\Lambda-\Lambda$ comparison, the $\Delta^0 -\Delta^{++}$ pair
 does not suffer from large difference in statistics and misidentifications,
 associated with rare $\bar\Lambda$ species.
 
 Since $\Delta$ resonance has been studied by STAR collaboration previously, by identifyng
 invariant mass peak on top of combinatorial background, what remains to be done
 is to correlate the decay vectors with the event plane. To avoid mixing with the directed flows of $\Delta$,
the decay direction should of course be determined in its rest frame.

 {\bf Acknowledgements.}
The idea of this comment came during the talk of K.Fukushima at CPOD 2016, who discussed there
different effects of the magnetic fields near freezout. I also would like to thank
Zhangbu Xu for discussion.
This work was supported in part by the U.S. D.O.E. Office of Science,  under Contract No. DE-FG-88ER40388.


\end{document}